# Einstein's Electron and Local Unitary Branching: Boundaries of Islands of Coherence and Quantum Nonlocality

Xing M. Wang[1]

## Abstract

The Branched Hilbert Subspace Interpretation (BHSI) aims to provide a unitary account of quantum measurement while maintaining a single-world ontology. The framework reexamines scenarios such as Einstein's 1927 electron-diffraction thought experiment by treating measurement as a finite dynamical process of information recording, comprising a sequence of unitary operations: branching, engaging, and disengaging. This perspective motivates a testable proposal: a dual-layer experiment in which the particle transit time between layers is shorter than the sensor response time, enabling a direct probe of measurement timing and potentially "uncommitted" outcomes. We introduce the Island of Coherence (IOC) as an operationally isolated quantum system, mathematically described by a Local Hilbert Subspace (LHS), which coexists with the background spacetime and within which unitary branching occurs. Historically, the first quantization already implies this dual structure. Applying the Gleason and Busch theorems to local unitary branching, the Born rule follows from the amplitudes given in the initial state. Moreover, quantum nonlocality (e.g., in Bell tests or tunneling) arises naturally from the inner-product structure of the LHS, which possesses no intrinsic spacetime metric. BHSI thus provides a coherent framework in which relativistic causality and quantum correlations remain structurally compatible.

## 1. Introduction

In our previous article [1], we proposed the Branched Hilbert Subspace Interpretation (BHSI). In this framework, measurement is modeled as a sequence of unitary operators: branching, engaging, and disengaging within the local Hilbert space (LHS). The resulting branches are locally decoherent, evolve unitarily and independently, and the system's initial state determines their amplitudes—thereby encoding the Born rule. Notably, branches may recohere before irreversibly entangling with the environment, as explored in detail in [2]. We have demonstrated that such locally controlled decoherent-recoherent processes are observable in protocols such as quantum teleportation ([3]; Sec. 5.2 of [1]). Building on this, we proposed experiments using modern Stern-Gerlach interferometers (SGI) [4,5] to visualize the physical reality of branch weights, branch-dependent electromagnetic and gravitational phase shifts ([4,5], §5.3 of [1]), and to probe the potential for recoherence using dual-sensing SGI [2].

[1] Sherman Visual Lab, Sunnyvale, CA 94085, USA; xmwang@shermanlab.com; ORCID:0000-0001-8673-925X

To directly probe local quantum branching, let us revisit Einstein's famous thought experiment presented at the 1927 Solvay Conference [6,7]. The experiment involved a screen with a small opening, through which electrons (or photons) were directed. Behind this screen was a large hemispherical photographic film to record the particles' landing positions. Quantum theory describes particles as waves (de Broglie waves). These waves diffract at the opening, resulting in a distribution of particle detections on the film. According to the Copenhagen Interpretation (CI, [8, 9]), when an electron reaches a specific position on the film, it suddenly finds itself at that particular location, and the probability of finding it elsewhere vanishes simultaneously (a collapse). Einstein argues: "The interpretation, according to which [the square of the wave-function] expresses the probability that this particle is found at a given point, assumes an entirely *peculiar* mechanism of action at a distance, which prevents the wave continuously distributed in space from producing an action in two places on the screen." In 1927, Einstein called this action at a distance "peculiar," not "spooky, but they were referring to the same concept.

Einstein's thought experiment provides an ideal scenario for comparison between the Many-Worlds Interpretation (MWI, [10-12]), CI, and the BHSI. The setup of Einstein's thought experiment is now achievable with modern single-electron sources [13], sub-nanometer to few-nanometer scale pinhole [14,15], and nanosecond-resolution detector arrays. In the next section, we describe how to realize the experiment, ensuring that no signals propagate outside the closed system of the opaque hemisphere of modern opaque electron sensors [16,17]. In Section 3, we describe the process mathematically and compare the unitary branching of MWI with that of BHSI. In addition, the uneven distribution pattern can be recorded in advance using a scintillating screen and an external optical camera [13, 18] to visualize the Born rule.

To further probe the dynamics of local branching, in Section 4, we propose a novel dual-layer detector system featuring a transparent inner hemisphere with transparent electron sensors [19, 20], aligned with opaque sensors on the outer hemisphere (dual-sensing). The electron's transit time between layers is comparable to the reaction times of modern sensors. This dual-layer design allows us to investigate scenarios of potential misaligned detections (e.g., inner #35 → outer #45), which would offer profound insights into the dynamics of quantum branching. In Section 5, we analyze possible outcomes (normal or abnormal) of the dual-layer experiment, distinguishing among interpretations. The setup ensures that the measurement process (branching, engagement, and disengagement) occurs entirely within the measured system.

In Section 6, we formalize this viewpoint by introducing the concept of the *island of coherence* (IOC), which is an operationally bounded quantum system that behaves as a coherent, inseparable whole under measurement and is described by a local Hilbert space (LHS). Within such an LHS, the Born rule can be derived via the Gleason and Butch theorems [21-22]. Moreover, quantum states are *intrinsically nonlocal*, since a Hilbert space is a vector space equipped with an inner product but without a spacetime metric [23]. This intrinsic nonlocality gives rise to interference, tunneling, and Bell-type correlations without violating relativistic causality. Consequently, branching occurs only within the LHS of the measured system and does not extend to other systems that are not operationally entangled with it. Moreover, the first quantization [9] indicates that the LHS of an IOC *coexists* with the spacetime region in which the IOC physically resides, making relativistic causality and quantum correlations comparable.

This leads to a powerful and parsimonious picture: each quantum system under observation is an island of coherence (IOC) inseparable by observation, surrounded by an effectively classical environment. Its correlations with other quantum "islands" are negligible unless they are deliberately entangled. Such islands may range from a pair of entangled photons to billions of Cooper pairs involved in macroscopic superconducting tunneling and, in principle, to astronomical objects such as white dwarfs and neutron stars [24-31].

By clarifying the roles of boundaries of IOCs and the intrinsic nonlocality of LHSs, BHSI resolves the measurement problem in a way that preserves unitarity, aligns with laboratory practice and decoherence theory, and avoids the wavefunction collapse in CI, the ontological excess of MWI, and the explicit nonlocality of Bohmian mechanics [32-33].

## 2. Localizing Quantum Branching: Single-Layer Hemispheric Detector

The core experimental setup aims to realize a modern version of Einstein's thought experiment on electron diffraction, directly probing the quantum branching in a local Hilbert space.

A highly collimated beam of single electrons, each with an energy of approximately 1 keV, is emitted from a controlled source at a low rate, e.g., $f$ ~ 1 MHz, ensuring 1 μs separation of individual electrons. This beam is directed through an exquisitely small pinhole, which induces significant diffraction of the electron's wave function. The diffracted electron then propagates towards a large, hemispherical detector array, positioned so that the pinhole effectively serves as the center of the sphere. This detector, with a radius $R$ ~ 10 cm, comprises 1000 individually addressable opaque sensors (reaction time τ ~ 0.1 ns), designed to register the arrival of a single electron (Fig. 1). The experiment focuses on recording the precise location (which sensor) and time of arrival for each electron.

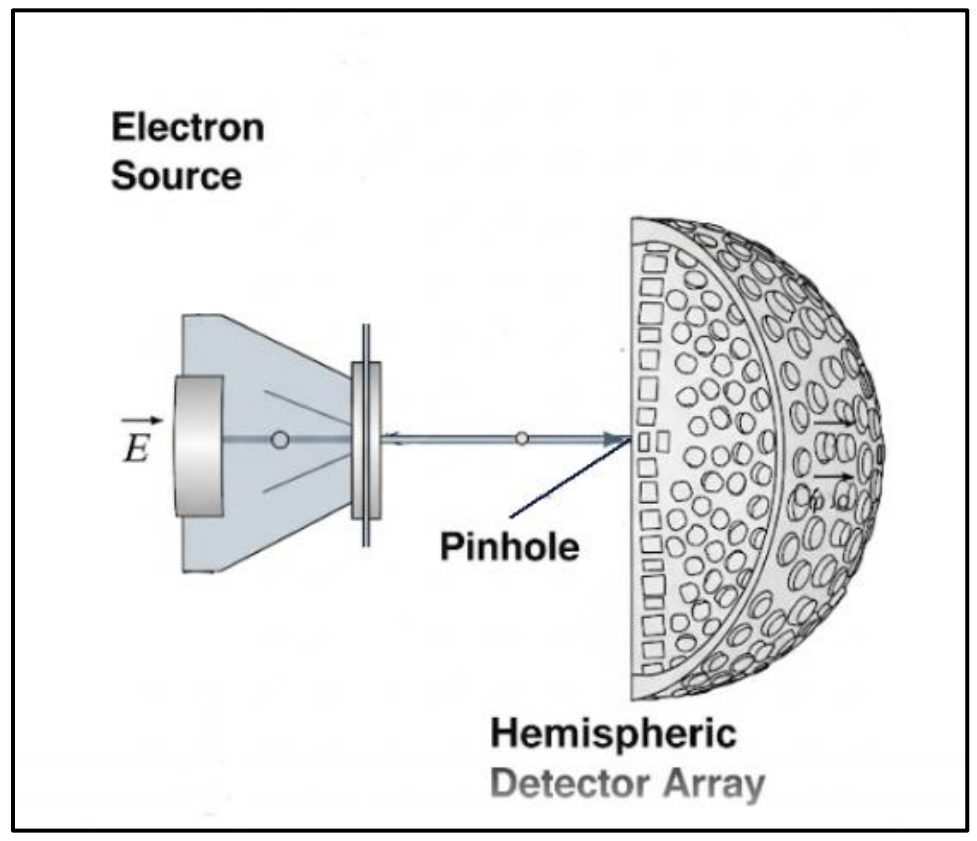

Fig. 1: Schematic Diagram of a Single-Layer Opaque Detector

The feasibility of constructing this primary experimental setup relies on significant advances in modern electron and detector technologies, which are conducted under tightly controlled environmental conditions. The entire experimental apparatus must operate under Ultra-High Vacuum (UHV) conditions, typically at a pressure below 10−9 Torr. This is essential to minimize electron scattering by residual gas molecules, which would otherwise obscure the delicate diffraction pattern. UHV also extends the cathode's lifespan and prevents contamination of the pinhole and detector surfaces. The experiment can generally be conducted at room temperature. However, precise temperature stabilization might be beneficial for long-term drift control, and some high-performance detectors may incorporate localized cooling to minimize noise. Key components and their current technological status include:

- **Single-Electron Source:** Stable, high-brightness single-electron emission is routinely achieved with advanced Field Emission Guns (FEGs), commonly found in modern electron microscopes. These sources provide highly coherent electron beams suitable for single-electron experiments [13].
- **Pinhole:** The challenging sub-nanometer to few-nanometer scale pinhole required for significant electron diffraction is at the cutting edge of nanofabrication, but is achievable. Techniques such as Focused Ion Beam (FIB) milling or advanced Electron Beam Lithography (EBL) can sculpt apertures with nanometer precision [14, 15].
- **Hemispheric Detector Array:** The core of the detection system, this array can be realized by precisely tiling numerous high-performance direct electron detectors (DEDs), typically based on CMOS or Hybrid Pixel Array (HPA) technologies, onto a machined hemispheric support structure. These detectors offer single-electron sensitivity, rapid readout capabilities (with a reaction time of $\tau \sim 0.1$ ns and high frame rates), and high quantum efficiency with minimal dead space, making them ideal for single-electron counting experiments [16, 17].

## 3. Mathematical Description and Interpretations of the Procedure

**The Initial State**: When an electron is emitted through the pinhole, its wave propagates in the hemisphere. Because we only concern ourselves with the events in which the electron is detected by one of the $N = 1000$ sensors, the total wave function can be written in two parts:

$$|\Psi\rangle = a|\Psi_e\rangle + b|\Psi'\rangle, \quad |\Psi_e\rangle = \sum_{k=1}^{N} c_k |\psi_k\rangle, \quad \sum_{k=1}^{N} |c_k|^2 = 1, \quad \prod_{k=1}^{N} |c_k|^2 \neq 0, \quad N = 1000 \tag{1}$$

Here, the wave part $|\Psi_e\rangle$ represents the normalized initial state, a superposition of $N$ possible outcomes with non-zero probabilities, as described by Eq. (1) of [1]; the wave part $|\Psi'\rangle$ represents any undetected electron events that occur when electrons are caught in the area between the sensors or on the bottom. The basis states of the initial state can be considered as the eigenstates of the operator of the sensor's serial number:

$$\hat{n}|\psi_k\rangle = k|\psi_k\rangle, \quad \langle\psi_i|\psi_k\rangle = \delta_{i,k}, \quad i,k \in \{1,2,\ldots,N\}, \quad N = 1000 \tag{2}$$

**3.1. BHSI Interpretation**: The existence of $|\Psi\rangle$ means the quantum system is described by an inseparable wavefunction in a *single LHS.* The whole process can be described as follows.

**The Branching**: When the wave front touches the hemisphere ($R$ ~ 10 cm), it starts the first operation of the measurement process, branching, as described by Eqs. (2-3) in [1]:

$$\hat{B}(|\Psi_e\rangle\otimes|E\rangle_L)\equiv|\Psi_B\rangle=\sum_{k=1}^{N}c_k\,|\psi_k\rangle\,|E_k\rangle_L=\sum_{k=1}^{N}c_k\,|\psi_{B;k}\rangle,\quad |\psi_{B;k}\rangle\equiv|\psi_k\rangle\,|E_k\rangle_L \tag{3}$$

**The engaging-disengaging process**: Assume sensor #35 registers a hit, the engaging-disengaging process is $\Sigma_\beta \equiv \Gamma_\beta \mathrm{T}_\beta \Lambda_\beta$, as outlined in Eqs. (4-6) in [1]:

$$|\Psi_B\rangle\otimes|\text{ready}\rangle_O\rightarrow\sum\nolimits_{k=1}^{1000}c_k(1-\delta_{k,35})|\psi_{k,B}\rangle+c_{35}\,|\psi_{35,B}\rangle\,|\text{reads }35\rangle_O\rightarrow|\Psi_B\rangle\otimes|\text{ready}\rangle_O \tag{4}$$

**The relocating process**: Because the electron is detected and absorbed by the sensor, there is *zero probability of finding the electron anywhere outside the hemisphere*. Therefore, the decoherent branches must have been entangled within the closed local environment:

$$U_E:|\Psi_B\rangle\otimes|E\rangle\rightarrow|E'\rangle \tag{5}$$

Otherwise, the conservation of energy, matter, and charge is violated.

### 3.2. Many-Worlds Interpretation (MWI):

According to MWI, when the electron interacts with the detector, the universe instantly branches into 1,000 decoherent worlds — one for each sensor that could have recorded the detection. Suppose sensor #35 in our world registers the electron. Then, in the 999 parallel but causally disconnected branches, the electron is detected by other sensors. The observer who reads "35" is simply the version of the original observer that became entangled with the sensor #35 outcome. Decoherence — the entanglement of the sensor with the global environment — prevents the branches from interfering with each other.

$$|\Psi_e\rangle\otimes|E\rangle_G\rightarrow\sum\nolimits_{k=1}^{1000}c_k\,|\psi_k\rangle\,|E_k\rangle_G,\quad {}_G\langle E_j\,|\,E_k\rangle_G\rightarrow\delta_{j,k} \tag{6}$$

The rate of generating new worlds is about one billion per second in our lab setting. Notably, since each branch contains a complete and closed copy of the experimental system, the observed electron should leave no detectable signal beyond the hemispherical detector, preserving conservation of energy, matter, and charge in every branch.

### 3.3. Copenhagen Interpretation (CI):

In CI, when the electron interacts with sensor #35, the entire wavefunction — previously spread over all sensors — undergoes an instantaneous, non-unitary collapse to a single point. The probability of detecting the electron in the other 999 sensors instantly drops to zero. This collapse occurs at the moment of measurement, without a detailed account of its physical mechanism. Since the collapse reduces a spread-out wave into a sharply localized result instantaneously, it implies a form of nonlocality — the kind Einstein famously criticized as "spooky action at a distance."

### 3.4. Comparing Interpretations:

Both MWI and BHSI maintain unitary evolution and avoid the postulated collapse of CI.

However, they differ fundamentally in ontology. MWI asserts that all possible outcomes occur in parallel, real worlds, each branching irreversibly upon detection. In contrast, BHSI posits a branching structure of local Hilbert subspaces in a single World. Alternative branches exist temporarily but become inaccessible as they rapidly entangle with the environment. In this view, branching is real, but it occurs within a single Universe and does not generate new, unobservable realities. In addition, because LHS's intrinsic nonlocality and its coexistence with the background spacetime, there is no spooky action in BHSI's view (§6.2)

### 3.5. The Born Rule:

The diffraction pattern need not be spherically symmetric; it can be shaped by adjusting the pinhole's size and geometry. In such cases, we can pre-record the expected intensity distribution optically by using a scintillating screen placed on the inner surface of a hemisphere (radius $R$ about 10 cm, without sensors), paired with an external scientific camera (e.g., CMOS or CCD) [13,18]. This optical map, representing the probability density $|\Psi|^2$ according to the Born rule, provides a reference distribution against which we compare the detector click statistics. When the experiment is repeated with sensors on the opaque hemispheric detector, the observed click distribution, compiled over billions of emitted electrons, should match the pre-recorded intensity profile. In BHSI, this agreement arises naturally: the branch weights, defined by the squared amplitudes in each local Hilbert subspace, $|c_k|^2$, obey the Born rule and govern the statistical frequencies of detector outcomes (see the derivation in §6.1.3).

## 4. Enhanced Experimental Setup: Dual-Layer Hemispheric Detector

To rigorously probe the subtle dynamics and completeness of the quantum measurement process, we propose an advanced dual-layer detector system (dual-sensing). This setup builds upon the single-layer experiment by introducing a second, inner hemispheric detector array that is critically transparent to the incoming electrons. The electron beam, after diffraction through the pinhole, first encounters the inner transparent detector, positioned at a radius of $R$ ~ 19.5 cm. This layer comprises numerous (e.g., 200) individually addressable segments or sensors. After interacting with the inner layer, the electron continues its trajectory a very short distance, approximately 0.5 cm, to the 200 sensors at the outer opaque hemispheric detector ($R$ ~ 20 cm), which is similar in design to the detector in the primary experiment. Both layers are precisely aligned (Fig. 2). The experiment's core measurement involves recording "double-click" events: correlated detections in both inner and outer layers within an extremely narrow time window.

The extremely short physical separation of $\Delta R$ ~ 0.5 cm between the detector layers, combined with the higher electron energy of 5 keV (v ~ $4.2 \times 10^7$ m/s), yields a remarkably short electron transit time of $\Delta t \sim \Delta R/\mathrm{v} \sim 0.12$ ns. This timescale is critical, as it is comparable to, or potentially even shorter than, the full 'reaction' or 'decision' time of the fastest modern transparent single-electron detectors ($\tau_{\mathrm{in}}$ ~ 1 ns). This unique temporal window enables the experiment to investigate whether the quantum measurement event initiated at the inner transparent detector is truly instantaneous and irreversible, or a dynamic process (a time-like sequence of space-time events) that takes a finite amount of time to complete.

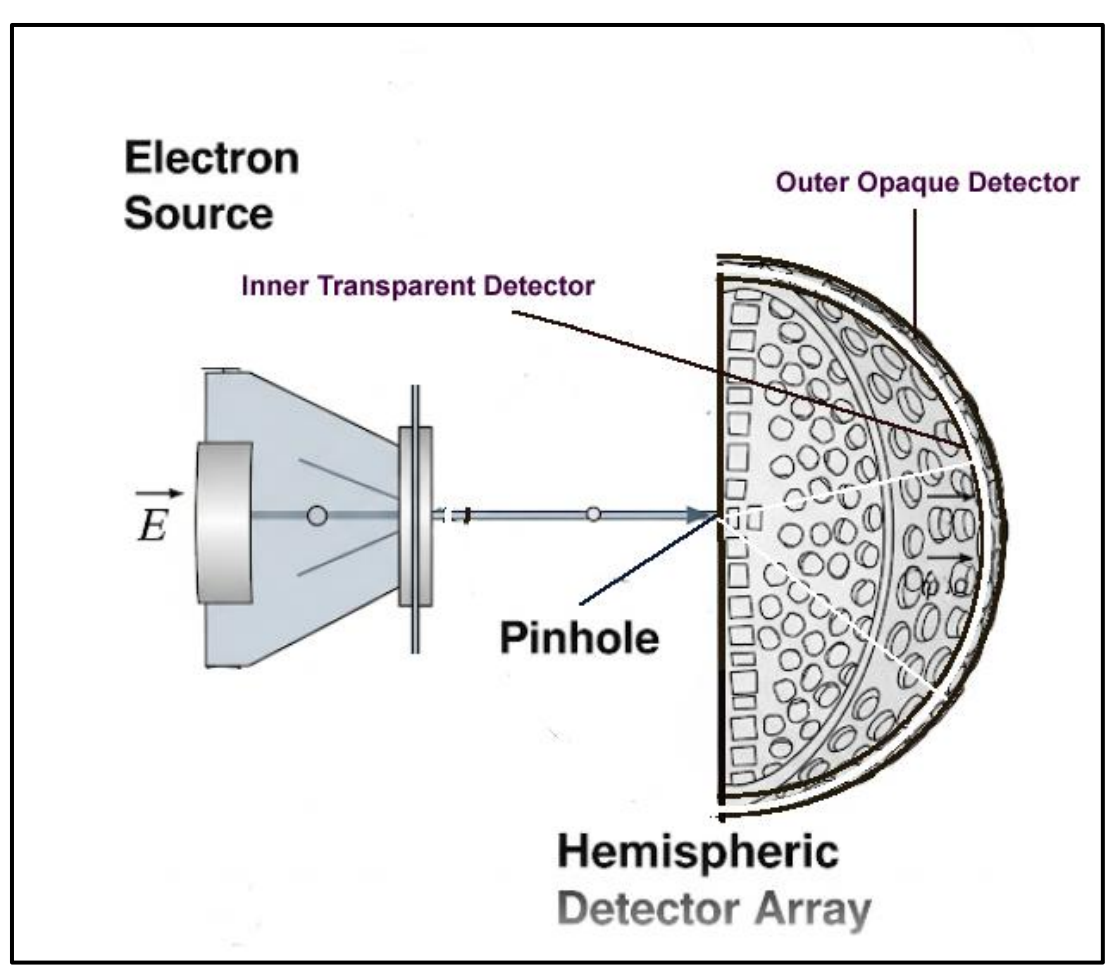

Fig. 2: Schematic Diagram of a Dual-Layer Hemispheric Detector

The electron source, operating at 5 keV, remains a standard Field Emission Gun (FEG) system capable of generating individual electrons in Ultra-High Vacuum (UHV) environments [13]. Similarly, the demanding nanometer-scale pinhole can be achieved through advanced FIB milling techniques [14]. The outer opaque hemispheric detector, composed of tiled direct electron detector (DED) modules (e.g., CMOS or hybrid pixel arrays), is also within current manufacturing capabilities, ensuring robust single-electron detection with high efficiency and precise timing (reaction time $\tau \sim 0.1$ ns).

The paramount technological challenge for this enhanced setup lies in the inner transparent hemispheric detector array. This component requires materials that are exceptionally thin yet robust (e.g., graphene, ultrathin silicon nitride, or amorphous carbon membranes) to minimize electron scattering and energy loss, ensuring that electrons propagate to the outer layer. Simultaneously, these transparent segments must be active detectors, capable of generating a measurable signal from a single 5 keV electron with a reaction time $\tau_{in} \sim 1$ ns, providing precise sub-nanosecond timing, and maintaining spatial addressability across hundreds of segments. Integrating active detection elements (such as highly sensitive 2D-material-based sensors or ultra-thin silicon structures) onto such large, curved, transparent substrates, while maintaining minimal interaction with passing electrons and providing rapid readout, represents the state of the art in current detector research and fabrication science [19, 20]. While extremely ambitious, this component conceptually aligns with ongoing efforts in novel electron microscopy detectors and atomically thin material-based sensing, framing the dual-layer experiment as a powerful, aspirational grand challenge for foundational quantum physics.

The time window $T_W$ (~6 ns) for counting any two successive clicks is set as follows:

$$\Delta t\,(\sim \Delta R\,/\,\mathrm{v} \sim 0.12\ \mathrm{ns}) \le \tau_{in}\,(\sim 1\,\mathrm{ns}) < T_W\,(\sim 6\,\mathrm{ns}) << 1/f\,(\sim 1\ \mu\mathrm{s}) \tag{7}$$

Because the transit time $\Delta t$ (~0.12 ns), plus the reaction time ($\tau$ ~0.1 ns) of the outer sensors, is approximately or shorter than the inner sensor reaction time ($\tau_{in}$ ~1 ns), we can expect possible "*delayed choices*" or "*uncommitted choice*" by the inner sensors during the test runs within the time window ($T_W \sim 6$ ns, which includes $\pm 3\sigma \sim \pm 3$ns, the standard deviations).

We can also pre-record the wave distribution density on a scintillating screen placed on the inner surface of the outer hemisphere (radius $R$~20 cm, without the inner detector layer) to visualize the branch weights, as already described in Section 3.4.

## 5. Interpretations of Possible Two-Layer Experimental Results

This two-layer detection setup allows us to probe when and how measurement-induced branching occurs by analyzing correlated detection events across three major categories.

**5.1**: **Aligned Detection**: Inner Sensor #35 → Outer Sensor #35.

This is the expected and dominant outcome: an electron passes through transparent sensor #35 in the inner layer. It is subsequently propagated to and absorbed by sensor #35, which is aligned in the outer layer. The two events are separated by a consistent time delay (within the ~6 ns window), confirming that they represent the same particle.

**BHSI:** Branching occurs locally at the inner sensor, where the electron's wavefunction decoheres into 200 branches. One branch engages/disengages with sensor #35 with the probability $|c_{35}|^2$, and propagates to the outer sensor, deterministic and unitary, as described by Case 1 in Section 2.2 of [1], for the basis state $|\psi_{35}\rangle$. All 200 branches, $|\psi_{k,\,B}\rangle$, are relocated to the environment before exiting the outer layer. This preserves a single-world ontology while explaining the Born rule naturally from the amplitudes of the initial state.

**MWI:** The electron evolves into a superposition of all possible paths, each corresponding to a different world. The observer experiences one outcome (e.g., #35 → #35), while the other 199 outcomes exist in parallel but inaccessible worlds. This is consistent with MWI but leaves the ontology bloated and unverifiable in practice.

**CI:** The wavefunction collapses instantaneously at the inner detector (#35), and the particle is then treated classically en route to the outer detector. However, the fact that all other inner sensor probabilities drop to zero instantaneously raises Einstein's concern about "spooky action at a distance."

**5.2: Misaligned Detection**: Inner Sensor #35 → Outer Sensor #45.

Rare but possible outcomes, where the inner sensor #35 fires, but the outer detection occurs at a different location (e.g., #45), within the timing window (~ 0.6 ns). They may reflect a slight scattering in the transparent layer or detector misalignment. Otherwise:

- **BHSI:** It may imply that the branch $|\psi_{45,\,B}\rangle$ arrives at the outer layer before the inner sensor #35 has registered (uncommitted outcome). Since BHSI models measurement as a sequence of local, unitary events (branching, engaging, disengaging, relocalizing), such anomalies are fully compatible with, and potentially informative for, estimates of the actual timescales of these events.
- **MWI:** MWI assumes global, timeless branching. A mismatch in real-time detection challenges this view, especially if the inner sensor registers first. MWI lacks a mechanism to explain how a superposition "chooses" a mismatched outer outcome in a given branch without violating its principle of simultaneity.

- **CI:** This result directly contradicts the idea of instantaneous wavefunction collapse at the inner sensor. If the particle's position was fixed at #35, why would it appear at #45 later?

**5.3: One Outer Detection Only**. No inner sensing → Outer Sensor #45.

In this scenario, the outer opaque detector registers the electron, but the corresponding inner transparent sensor fails to record a signal. While often attributed to detector inefficiency, the consistent statistical occurrence of these anomalous events suggests a "temporal threshold effect" unique to BHSI. However, whether global probability conservation can be temporarily violated in the fuzzy boundary zone remains an open question.

The diagnostic power of this dual-layer setup lies in its ability to create a clean empirical testbed. It allows us to distinguish between interpretations by directly probing the time-resolved and local nature of quantum measurement. Even if most outcomes are consistent across multiple interpretations, subtle anomalies (Sections **5.2** or **5.3**) could provide discriminating empirical evidence in favor of a local Hilbert space within a single-world ontology, as proposed by BHSI.

Note: Scenarios involving apparent violations of energy or particle conservation (e.g., duplicate detections on inner or outer layer only) are excluded from consideration, as such events would indicate experimental artifacts or a breakdown of quantum theory itself rather than differences between interpretations.

## 6. The Boundary and Nonlocality of Islands of Coherence

Throughout this paper, we emphasize that quantum measurements occur within islands of coherence (IOC). It is therefore essential to clarify what is meant by an "IOC," why such islands exist, and how their intrinsic properties allow for unitary branching without necessitating a collapse or global "Many-Worlds" split.

### 6.1: Island Ontology, Boundary, and the Born Rule

An Island of Coherence (IOC) physically represents a finite, dynamically inseparable quantum system that maintains its coherent quantum state in the presence of negligible environmental decoherence. An IOC is mathematically described by a local Hilbert space (LHS).

**6.1.1. The operationally isolated IOC**: An IOC is a system—microscopic, mesoscopic, or macroscopic—that is *operationally* isolated from its environment. This isolation is defined by a *quantum boundary*: a threshold at which the correlation (decoherence) between the system ($Q_0$) and its environment ($E$) is sufficiently low—as defined by the relevant measurement context—to permit the system's coherent unitary evolution. This boundary separates the system from surrounding classical or quantum systems, effectively defining a local tensor-product factorization of the total state space, in which $Q_0$ does not form an inseparable quantum whole with $E$, as in Eq. (8) below.

A hypothetical universal ("World") wavefunction $|W\rangle$ may be formally defined in principle, but BHSI treats it as operationally separative. Observable physics is instead described in terms of separable classical (C) systems and inseparable quantum (Q) systems, each associated with its own operational domain. When $Q_0$ is measured, the dynamics are strictly confined to $Q_0$'s local

Hilbert space, effectively partitioning the World state and its probability density into functionally independent domains:

$$\begin{aligned} &|\,\mathrm{W}\rangle \rightarrow \bigotimes_i |\,C_i\rangle \bigotimes_\mu |\,Q_\mu\rangle \rightarrow |\,Q_0\rangle; \\ &\rho(\mathrm{W}) \rightarrow \bigotimes_i \rho(C_i) \bigotimes_\mu \rho(Q_\mu) \rightarrow \mathrm{Tr}_E \rho(\mathrm{W}) \rightarrow \rho(Q_0), \quad \text{where } Q_0 \notin E \end{aligned} \tag{8}$$

Here, the indices $i$ label classical subsystems, $\mu$ label quantum subsystems, and $E$ denotes all degrees of freedom external to the IOC of $Q_0$. The final expressions represent the effective quantum state relevant to the measurement, obtained by tracing over inaccessible environmental degrees of freedom and by renormalization.

The existence of IOCs is the consequence of the conditions required for quantum measurement: any quantum experiment requires isolating the system of interest from uncontrolled environmental degrees of freedom to a sufficient degree. The boundaries of these islands are not arbitrary; they are dynamically maintained by environmental decoherence, which acts as the dissipative "current" of the classical ocean, suppressing interference between the system and its surroundings.

**6.1.2. The Fuzzy Boundary:** For each island of coherence, there is an outer boundary, beyond which degrees of freedom are considered decohered and separable from the quantum system, and an inner boundary, within which all components evolve as a single unitary whole. These two boundaries are generally not coincident; there is no sharp, instantaneous cut between the classical world and a quantum island of coherence. This fuzziness of the quantum-classical boundary, a semi-classical spatiotemporal zone, is one of the core features of BHSI, arising from its view of local Hilbert spaces. For example, in the two electron-diffraction setups, the outer boundary is the OD layer's outer edge. In contrast, the inner boundary is the inner edge of the OD layer in the single-layer setup and the inner TS layer in the dual-layer setup.

**6.1.3. The Born Rule.** Since branching occurs within a single LHS ($\mathcal{H}_L$), each branch retains the amplitude assigned by the initial state vector. The probability of registering a given branch (the weight of the branch) is therefore determined by the structure of probability measures on projection operators within that LHS. By the Gleason theorem [21], for dim $\mathcal{H}_L \geq 3$, any non-contextual probability measure must take the Born form. For two-dimensional LHSs, the result follows from the Busch theorem [22]. Thus, the Born rule is derived from the internal Hilbert-space structure of a single IOC, without collapse or the coexistence of many worlds.

**6.2: The Intrinsic Nonlocality of a Local Hilbert Space and the Dual Structure**

A Hilbert space (or its rigged extension for continuous spectra [23]) is fundamentally a vector space equipped with an inner product. Crucially, it lacks a spacetime metric. Quantum states are vectors characterized by direction and magnitude within this vector space. This has a direct and profound consequence for spatial description. In the coordinate representation, distinct spatial positions $\boldsymbol{x}$ and $\boldsymbol{x}'$ ($\boldsymbol{x} \neq \boldsymbol{x}'$) correspond to two orthogonal basis vectors $|\boldsymbol{x}\rangle$ and $|\boldsymbol{x}'\rangle$, regardless of their physical distance $|\,\boldsymbol{x} - \boldsymbol{x}'|$:

$$\text{Hilbert Space: } \langle \boldsymbol{x} \mid \boldsymbol{x}' \rangle = 0 \text{ for } \boldsymbol{x} \neq \boldsymbol{x}' \Leftrightarrow \text{Euclidean Space: } \mid \boldsymbol{x} - \boldsymbol{x}' \mid > 0 \text{ for } \boldsymbol{x} \neq \boldsymbol{x}' \tag{9}$$

Likewise, the wave function amplitudes $\langle x|\Psi\rangle = \Psi(x)$ and $\langle x'|\Psi\rangle = \Psi(x')$ are not separated by any notion of physical distance in the Hilbert space; they are different components of the same state vector expressed in a chosen basis.

**6.2.1. The metric of a Hilbert space**: Although the Hilbert space has a mathematical metric induced by its inner product, it has no physically meaningful geometric distance. For example, the square of the distance between two distinct orthonormal basis vectors in Eq. (1), given by the metric defined in Hilbert space, is a constant, invariant of the spatial distance:

$$d_{i,j}^2 \equiv \langle \psi_i - \psi_j \mid \psi_i - \psi_j \rangle = \langle \psi_i \mid \psi_i \rangle + \langle \psi_j \mid \psi_j \rangle = 2 \quad \text{for } \forall i \neq j, \ \because \langle \psi_i \mid \psi_i \rangle = \delta_{i,j} \tag{10}$$

This intrinsic, non-spatial nonlocality is the foundational mechanism behind quantum tunneling, interference, and, critically, the nonlocal correlations observed in entangled systems [22-29]. These phenomena do not imply superluminal signaling or violate relativistic causality, because "influence" within a Hilbert space is not a causal process propagating through spacetime; it is a mathematical expression of the state's coherence.

If two systems are prepared in an entangled state and their coherence is preserved, they constitute a single IOC, described by a single LHS, irrespective of their spatial separation. This is precisely the case in a Bell-type experiment [22, p. 25]. A source generates a pair of photons in an entangled polarization state, such as one of the Bell states:

$$|\Phi^{\pm}\rangle = \frac{1}{\sqrt{2}}(|H\rangle_A |H\rangle_B \pm |V\rangle_A |V\rangle_B), \ |\Psi^{\pm}\rangle = \frac{1}{\sqrt{2}}(|H\rangle_A |V\rangle_B \pm |V\rangle_A |H\rangle_B) \tag{11}$$

This state does not reside in two separate 2D Hilbert spaces for photons *A* and *B*; it is an inseparable vector in the 4D tensor-product space $\mathcal{H}_A \otimes \mathcal{H}_B$, which forms the LHS for this composite system. If the photons are spatially separated while maintaining entanglement (i.e., without decohering into a separable mixture), they continue to occupy this single LHS, even if they are a light year apart. Therefore, synchronized measurements on A and B are not independent operations: both act on the inseparable composite system. This explains the observed nonlocal correlations as a consequence of the *intrinsic nonlocality* of the LHS, rather than as "spooky action at a distance" through spacetime [6,7] or as guidance by a nonlocal pilot-wave function (Bohmian mechanics) [30,31].

**6.2.2. The Coexistence of Two Structures within an IOC:** An Island of Coherence (IOC) connects two distinct but coexisting structures: the spacetime embedding of a localized physical system and the Hilbert-space structure that supports its coherent quantum state. Spacetime carries a classical metric and relativistic causal structure, whereas Hilbert space provides the linear structure governing superposition and entanglement. These structures are conceptually independent: Hilbert space has no intrinsic spacetime metric, and spacetime geometry does not define Hilbert-space distance, which is based on the inner product, as Eq (10) demonstrates.

These two structures coexist rather than replace one another. For example: a particle confined in a box has its state represented by an LHS, while the box's spacetime region remains operationally classical; in a superconductor, the Cooper-pair condensate is described by an LHS, whereas the macroscopic wire itself is embedded in an effectively classical spacetime; in our electron diffraction experiments, the electron wavefunction resides in an LHS, while the detector region is described by classical geometry.

Historically, this coexistence is already implicit in the first quantization (p. 92, Ref. [9]). The substitution of 4-momentum by operator $p_\mu \rightarrow i\hbar\partial/\partial x^\mu$ ($\mu = 0,1,2,3$) replaces the classical description of a particle's motion in spacetime with the evolution of a wavefunction in Hilbert space, while the coordinates $x^\mu$ serve as representation variables of the wavefunction $\psi(x^\mu)$, rather than specifying a definite particle trajectory in the manifold. For example, in the non-relativistic case, the spacetime variables and Hilbert space are structurally unified in the Schrödinger equation:

$$E = H(\vec{x},\vec{p},t) = \frac{\vec{p}^2}{2m} + V(\vec{x},t) \underset{\text{quantization}}{\overset{\text{first}}{\longrightarrow}} i\hbar\frac{\partial}{\partial t}\psi(\vec{x},t) = \left\{-\frac{\hbar^2}{2m}\nabla^2 + V(\vec{x},t)\right\}\psi(\vec{x},t) \tag{12}$$

Thus, the IOC occupies spacetime as an inseparable quantum system, while its coherence structure resides in the associated LHS. Recognizing the coexistence of these two structures clarifies how relativistic causal locality in spacetime can remain compatible with nonlocal quantum correlations arising from Hilbert-space structure.

### 6.3: Mathematical Linking: Local Operators and Subspace Dynamics

The effectively bounded island of coherence (IOC), described by its local Hilbert space $\mathcal{H}_L$, provides the necessary "theater" for the BHSI operational sequence. In this interpretation, the transition from a quantum superposition to a definite classical record is not a global event but a sequence of unitary transformations of the state vector $|\Psi\rangle$ strictly within $\mathcal{H}_L$.

**6.3.1. Branching (*B*):** Within the $\mathcal{H}_L$ of the IOC, the branching operator ***B*** partitions the local state into decoherent subspaces (see Eq. (3) above and Eq. (3) of [1]):

$$\begin{aligned} &\hat{B}:|\Psi\rangle \rightarrow |\Psi_B\rangle = \sum_{k=1}^{D} c_k\,|g_k\rangle\,|E_k\rangle_L \equiv \sum_{k=1}^{D} c_k\,|g_{B;k}\rangle, \quad {}_L\langle E_j|E_k\rangle_L \approx \delta_{j,k} \\ &\hat{B}\left(\mathcal{H}_S \otimes \mathcal{H}_{L,E}\right) = \bigoplus_{k=1}^{D} \mathcal{H}_{S,k}(\text{span } \mathrm{c}_k\,|\,g_{B;k}\rangle), \quad |\langle g_k|\Psi\rangle|^2 = |\langle g_{B;k}|\Psi_B\rangle|^2 = |c_k|^2 \end{aligned} \tag{11}$$

Because the IOC is operationally isolated, branching ***B*** acts *only* within $\mathcal{H}_L$. The resulting state $|\Psi_B\rangle$ remains a single, inseparable vector in $\mathcal{H}_L$, but its components are now dynamically independent.

**6.3.2. Engaging ($\Lambda_\beta$) and Disengaging ($\Gamma_\beta$):** The observer's interaction with the branched system starts with a local engagement operator $\Lambda_\beta$ that correlates the observer's state with a specific branch $|g_\beta\rangle$ in accordance with the Born weights $|c_\beta|^2$ (see Eq. (4-5), [1]):

$$\Lambda_\beta : |\,\text{ready}\rangle_o \in \mathcal{H}_E \mapsto |\,\text{reads } g_\beta\rangle_o \in \mathcal{H}_{S,\beta}, \quad \beta \in \{1.2.\cdots D\} \tag{12}$$

Because the IOC is isolated, this engagement does not "drag" the rest of the universe into a split. After recording the outcome, operator $T_\beta$ changes the observer's state to |ready⟩, then operator $\Gamma_\beta$ disengages him from the branch. This operator effectively "seals" the IOC boundary, completing the measurement cycle. The branches, whose evolution is constrained by conservation laws, are eventually transferred unitarily into environmental degrees of freedom, becoming effectively suppressed through decoherence while maintaining unitary integrity (see Eq. (7, 24), [1]):

$$T_\beta : |\,\text{reads}\rangle_O \mapsto |\,\text{ready}\rangle_O; \quad \Gamma_\beta T_\beta : \mathcal{H}_B \mapsto \mathcal{H}_f = \left\{ \bigoplus_{k=1}^{D} \mathcal{H}_{S,k} (\text{span } c_k \,|\, g_{B,k}\rangle \right\} \otimes |\,\text{ready}\rangle_O \tag{13}$$

$$U_E : |\,\Psi_B\rangle \otimes |\,E\,\rangle \rightarrow |\,E'\,\rangle \tag{14}$$

Altogether, these operators provide a dynamical unitary alternative to "collapse." The total decoherence time is device-dependent, thereby defining the temporal extent of the fuzzy boundary zone. Throughout this process, no superluminal action occurs, even if the sensors are light-years apart: correlations are encoded in the intrinsic nonlocal structure of the LHS (§6.2).

### 6.4: Examples of Islands of Coherence

The Islands of Coherence (IOC) is not a mathematical abstraction. Its defining feature is the existence of a boundary—enforced by experimental isolation or intrinsic dynamics—within which the system must be treated as an inseparable quantum whole, as defined in Section 6.1. The electro-diffraction experiments analyzed in this paper are prime examples of such a system. This principle applies universally:

*Microscopic/Mesoscopic*: Entangled photon pairs in Bell-type experiments [24], single atoms undergoing quantum tunneling [25], and hybrid systems such as a photon entangled with a trapped ion in a high-finesse cavity [26] are all well-described by IOCs defined by experimentally maintained coherence and isolation. Examples of IOC in relativistic quantum field theory are discussed in §5.2 of [2].

*Macroscopic:* Superconductors and superfluids, such as the billions of electron Cooper pairs in macroscopic superconducting tunneling [27], where a single wave function describes a collective state. Notably, parallel quantum computing is more parsimoniously described as the evolution of parallel local Hilbert subspaces rather than the splitting of the device into parallel worlds [28].

*Astrophysical*: On much larger scales, compact astrophysical objects such as white dwarfs [26] and neutron stars [29] can be modeled, to good approximation, as quantum systems in which degeneracy pressure—a purely quantum effect—governs the macroscopic equilibrium of the entire object. Even black holes have been proposed to admit an effective description as condensates of soft gravitons in certain approaches to quantum gravity [30].

Many of the above examples are systems of multiple identical particles, such as the entangled photon pair mentioned in Bell tests, the Cooper pairs in macroscopic quantum tunnelling, and the neutrons in a neutron star. They are all identical particles, either fermions or bosons, forming an inseparable quantum island. No experiment can locate a specific particle or its trace without destroying the system's coherence.

## 7. Summary and Discussion

Four central features characterize the Branched Hilbert Subspace Interpretation (BHSI). First, a quantum system under measurement is described as an Island of Coherence (IOC): an operationally bounded quantum domain represented by a Local Hilbert Subspace (LHS); the LHS encodes correlations through its inner-product (vector-space) structure, which permits nonlocal entanglement, while the IOC coexists with and is embedded in the classical spacetime manifold, a dual structure implied by the first quantization [9]. Second, the IOC possesses an operation-dependent semiclassical boundary zone and a corresponding lifecycle [2] determined by its interaction with external degrees of freedom. Third, quantum measurement is treated as a finite, time-extended unitary process whose internal structure—branching, engagement, and disengagement—may become experimentally accessible as detection technologies advance. Fourth, local branching remains reversible in principle prior to irreversible environmental entanglement [1,2].

Revisiting Einstein's 1927 diffraction thought experiment [6,7], we analyzed a single-layer detector configuration in which unitary branching is confined within the LHS without invoking collapse or global wavefunction splitting. Within a single IOC, the Born rule can be derived from Hilbert-space structure and empirically observed. The proposed two-layer detector extends this analysis by introducing a transparent inner detection layer aligned with an outer absorbing array. This configuration is designed to probe the temporal structure of measurement. If branching is indeed a local, time-extended dynamical process [1], deviations from perfect inter-outer-layer correlation could reveal its internal sequence. Similar dual-sensing logic applies to the proposed Stern–Gerlach recoherence experiments [2].

BHSI maintains unitarity while providing a structured account of measurement dynamics within local Hilbert spaces of islands of coherence, avoiding wavefunction collapse, global splitting, or hidden variables. As experimental control improves, it may become possible to investigate measurement timing directly, allowing long-standing foundational questions to be addressed through controlled, potentially falsifiable experiments.

## Abbreviations

| | |
|---|---|
| BHSI | Branched Hilbert Subspace Interpretation |
| CI | Copenhagen Interpretation |
| IOC | Island of Coherence |
| LHS | Local Hilbert Space |
| MWI | Many-Worlds Interpretation |

## References

1. Wang, X. M. Quantum Measurement Without Collapse or Many Worlds: The Branched Hilbert Subspace Interpretation. *arXiv:2504.14791* (2025). https://doi.org/10.48550/arXiv.2504.14791
2. Wang, X. M. Stern-Gerlach Interferometers with Dual Sensing: Probing Recoherence and Lifecycles of Islands of Coherence. *arXiv:2508.16019* (2026). https://doi.org/10.48550/arXiv.2508.16019
3. Ursin, R., Jennewein, T., Aspelmeyer, M., et al. Quantum teleportation across the Danube. *Nature* **430**, 849–849 (2004). https://doi.org/10.1038/430849a
4. Margalit, Y., Dobkowski, O., Zhou, Z., et al. Realization of a complete Stern–Gerlach interferometer: Towards a test of quantum gravity. *Sci. Adv.* **7**, eabg2879 (2021). https://doi.org/10.1126/sciadv.abg2879
5. Bose, S., Mazumdar, A., Morley, G. W., et al. A spin entanglement witness for quantum gravity. *Phys. Rev. Lett.* **119**, 240401 (2017). https://doi.org/10.1103/PhysRevLett.119.240401
6. Bohr, N. General discussion at the Fifth Solvay Conference (1927). In: *Niels Bohr Collected Works*, Vol. **6**, pp. 99–106 (Elsevier, 1985).
7. Hossenfelder, S. What did Einstein mean by "spooky action at a distance"? Public lecture [transcript, YouTube] (2021).
8. Bohr, N. Can quantum-mechanical description of physical reality be considered complete? *Phys. Rev.* **48**, 696–702 (1935). https://doi.org/10.1103/PhysRev.48.696
9. Dirac, P. A. M. *The Principles of Quantum Mechanics*, 4th ed. (Oxford University Press, 1958).
10. Everett, H. "Relative state" formulation of quantum mechanics. *Rev. Mod. Phys.* **29**, 454–462 (1957).
11. Wallace, D. *The Emergent Multiverse* (Oxford University Press, 2012).
12. Vaidman, L. Why the many-worlds interpretation? *Quantum Rep.* **4**, 264–271 (2022). https://doi.org/10.3390/quantum4030019
13. Egerton, R. F. *Physical Principles of Electron Microscopy* (Springer, 2005).
14. Gadgil, V. J., Tong, H. D., Cesa, Y., Bennink, M. L. Fabrication of nanostructures in thin membranes with focused ion beam technology. *Surf. Coat. Technol.* **203**, 2436–2441 (2009). https://doi.org/10.1016/j.surfcoat.2009.02.036
15. Shahali, H., Hasan, J., Wang, H., et al. Evaluation of particle beam lithography for fabrication of metallic nanostructures. *Procedia Manuf.* **30**, 261–267 (2019). https://doi.org/10.1016/j.promfg.2019.02.038
16. Faruqi, A. R., McMullan, G., et al. Electron detectors for electron microscopy. *Microsc. Today* **24**, 34–41 (2016). https://doi.org/10.1017/S0033583511000035
17. Mendez, J. H., Mehrani, A., Randolph, P., Stagg, S. Throughput and resolution with a next-generation direct electron detector. *IUCrJ* **6**, 1007–1013 (2019). https://doi.org/10.1107/S2052252519012661

18. Holst, G. C., Lomheim, T. S. *CMOS/CCD Sensors and Camera Systems*, 2nd ed. (JCD Publishing, 2011).
19. Hassan, J. Z., Raza, A., Babar, Z. U. D., et al. 2D material-based sensing devices: An update. *J. Mater. Chem. A* **11**, 1–23 (2023). https://doi.org/10.1039/D2TA07653E
20. Ratti, L., Brogi, P., Collazuol, G., et al. Layered CMOS SPADs for low-noise detection of charged particles. *Front. Phys.* **8**, 607319 (2020). https://doi.org/10.3389/fphy.2020.607319
21. Gleason A. (1957). Measures on the Closed Subspaces of a Hilbert Space, *J. of Math. and Mech.*, **6**, 885-893. https://doi.org/10.1512/iumj.1957.6.56050. PDF Online.
22. Busch H. (1926). Calculation of the path of cathode rays in axially symmetric electromagnetic fields (in German). Ann. Phys. (Berlin) **386,** 974. https://doi.org/10.1002/andp.19263862507
23. Madrid, R. The role of the rigged Hilbert space in quantum mechanics. *Eur. J. Phys.* **26**, 287–312 (2005). https://doi.org/10.1088/0143-0807/26/2/008
24. Nielsen, M. A., Chuang, I. L. *Quantum Computation and Quantum Information*, 10th Anniversary ed. (Cambridge University Press, 2010).
25. Hartman, T. E. Tunneling of a wave packet. *J. Appl. Phys.* **33**, 3427–3433 (1962). https://doi.org/10.1063/1.1702424
26. Köbel, P., Breyer, M., Köhl, M. Deterministic spin–photon entanglement from a trapped ion in a fiber Fabry–Perot cavity. *npj Quantum Inf.* **7**, 6 (2021). https://doi.org/10.1038/s41534-020-00338-2
27. Martinis, J. M., Devoret, M. H., Clarke, J. Energy-level quantization in the zero-voltage state of a current-biased Josephson junction. *Phys. Rev. Lett.* **55**, 1543–1546 (1985). https://doi.org/10.1103/PhysRevLett.55.1543
28. Cuffaro, M. Many worlds, the cluster-state quantum computer, and the problem of the preferred basis. *Stud. Hist. Philos. Mod. Phys.* **43**, 35–42 (2012). https://doi.org/10.1016/j.shpsb.2011.11.007
29. Chandra, V., et al. Observation of long-theorized quantum phenomena in astrophysical systems. *Astrophys. J.* **899**, 146 (2020). https://doi.org/10.3847/1538-4357/aba8a2
30. Baym, G., Pethick, C., Pines, D. Superfluidity in neutron stars. *Nature* **224**, 673–674 (1969). https://doi.org/10.1038/224673a0
31. Ciliberto, G., Balbinot, R., Fabbri, A., Pavloff, N. Quantum backreaction in an analog black hole. *Phys. Rev. A* **112**, 063323 (2025). https://doi.org/10.1103/48sm-thzm
32. Bohm, D. A. Suggested Interpretation of the Quantum Theory in Terms of “Hidden” Variables. I & II. *Phys. Rev.* **85**, 166–193 (1952). https://doi.org/10.1103/PhysRev.85.166
33. de Broglie, L. *On the Theory of Quanta*. PhD Thesis, University of Paris (1924).